\documentclass[]{aa}
\usepackage{graphicx, natbib, longtable, lscape, amssymb}
\usepackage[english]{babel}
\usepackage{lscape}

\newcommand{\gppr}{\stackrel{>}{\scriptstyle \sim}}
\newcommand{\gappr}{\raisebox{-0.4ex}{$\gppr$}}
\newcommand{\lppr}{\stackrel{<}{\scriptstyle \sim}}
\newcommand{\lappr}{\raisebox{-0.4ex}{$\lppr$}}
\newcommand{\Mwd}{\mbox{$M_\mathrm{WD}$}}
\newcommand{\Msun}{\mbox{$M_{\odot}$}}
\newcommand{\Rsun}{\mbox{$R_{\odot}$}}
\newcommand{\Porb}{\mbox{$P_\mathrm{orb}$}}

\begin{document}

\title{Monte Carlo simulations of post-common-envelope white dwarf + main sequence binaries: The effects of including recombination energy}

\titlerunning{WD+MS PCEBs including recombination energy}
\author{M. Zorotovic\inst{1}, M.R. Schreiber\inst{1,2}, E. Garc\'{\i}a-Berro\inst{3,4}, J. Camacho\inst{3,4}, S. Torres\inst{3,4}, A. Rebassa-Mansergas\inst{5}, B.T. G\"ansicke\inst{6}}
\authorrunning{Zorotovic et al.}
\institute{$^1$Departamento de F\'isica y Astronom\'ia, Facultad de Ciencias, Universidad de Valpara\'iso, Valpara\'iso, Chile\\ 
\email{mzorotovic@dfa.uv.cl} \\
$^2$Millennium Nucleus ``Protoplanetary Disks in ALMA Early Science'', Universidad de Valpara\'iso, Casilla 36-D, Santiago, Chile\\
$^3$Departament de F\'isica Aplicada, Universitat Polit\`ecnica de Catalunya, c/Esteve Terrades 5, 08860 Castelldefels, Spain\\
$^4$Institute for Space  Studies of Catalonia, c/Gran Capit\`a 2--4, Edif. Nexus 104, 08034 Barcelona, Spain\\
$^5$Kavli Institute for Astronomy and Astrophysics, Peking University, Beijing 100871, China\\
$^6$Department of Physics, University of Warwick, Coventry CV4 7AL, UK
}
\offprints{M. Zorotovic}

\date{Received 13 November 2013 / Accepted 23 June 2014}

\abstract{Detached white dwarf + main sequence (WD+MS) post-common-envelope binaries (PCEBs) are perhaps the most suitable objects for testing predictions of close-compact binary-star evolution theories, in particular,
common-envelope (CE) evolution. Consequently, the population of WD+MS PCEBs has been simulated by several authors in the past and the predictions have been compared with the observations.
However, most of those theoretical predictions did not take the possible contributions to the envelope ejection from additional sources of energy into account (mostly recombination energy) stored in the envelope. }
{Here we update existing binary population models of WD+MS PCEBs by assuming that in addition to a fraction $\alpha_{\mathrm{CE}}$ of the
orbital energy, a fraction $\alpha_{\mathrm{rec}}$ of the recombination energy available within the envelope contributes to ejecting the envelope.} 
{We performed Monte Carlo simulations of $10^7$ MS+MS binaries for 9 different combinations of $\alpha_{\mathrm{CE}}$ and $\alpha_{\mathrm{rec}}$ using standard assumptions for the initial
primary mass function, binary separations, and initial-mass-ratio distribution and evolved these systems using the publicly available binary star evolution (BSE) code.}
{Including a fraction of the recombination energy leads to a clear prediction of a large number of long orbital period ($\gappr\,10$ days) systems mostly containing high-mass WDs.
The fraction of systems with He-core WD primaries ($\Mwd\lappr0.5\Msun$) increases with the CE efficiency and the existence of very low-mass He WDs ($\lappr\,0.3\,\Msun$)
is only predicted for high values of the CE efficiency, i.e. $\alpha_{\mathrm{CE}}\gappr\,0.5$. All models predict on average longer orbital periods for PCEBs containing C/O-core WDs
($\Mwd\gappr0.5\Msun$) than for PCEBs containing He WDs. This effect increases with increasing values of both efficiencies, i.e., $\alpha_{\mathrm{CE}}$ and $\alpha_{\mathrm{rec}}$. Longer
periods after the CE phase are also predicted for systems containing more massive secondary stars. The initial-mass-ratio distribution affects the distribution of orbital periods, especially the distribution of secondary star masses.} 
{Our simulations, in combination with a large and homogeneous observational sample, can provide constraints on the values ​​of $\alpha_{\mathrm{CE}}$ and $\alpha_{\mathrm{rec}}$, as well
as on the initial-mass-ratio distribution for MS+MS binary stars.}

\maketitle

\section{Introduction}\label{sec:intro}

Close binaries containing compact objects span a wide range of interesting and exotic stars, such as millisecond pulsars, galactic black hole candidates, detached white dwarf (WD) binaries, 
neutron star binaries, and interacting binaries, such as cataclysmic variables and low-mass X-ray binaries. The small binary separations of all these compact binaries imply that the radius of 
the progenitor of the compact object must have exceeded the current orbital separation quite far. How such close-compact binary systems could form was outlined more than 30 years ago by 
\citet{paczynski76-1}. The progenitors of close-compact binaries were initially relatively close binary systems ($a_{\mathrm{i}}\sim\,100-1000\Rsun$) consisting of two main-sequence (MS) stars. 
Once the primary, i.e. the more massive star, evolved off the MS and filled its Roche lobe during the first giant branch (FGB) or asymptotic giant branch (AGB), dynamically unstable mass transfer 
was generated, and the less massive star (from now on the secondary) could not accrete the transferred material, which thus started to accumulate around it and quickly formed a common envelope (CE); 
i.e., the envelope of the primary surrounded the core of the primary and the secondary star. Owing to drag forces between the envelope and the two stars, orbital energy was transferred from the binary 
(consisting of the core of the primary and the secondary) to the envelope, causing the binary separation to be significantly reduced and the CE to be ejected. After the envelope ejection, 
the system appears as a close but detached post-common-envelope binary (PCEB) consisting of a compact object, i.e. the core of the primary, and a MS star. 
Among the most numerous compact binaries are those containing a WD primary, and the stellar parameters are most easily measured if both stars are in a detached orbit. Such white dwarf + main sequence (WD+MS) 
PCEBs are therefore ideal systems for providing observational constraints on models of CE evolution.  

Binary population studies of PCEBs have been performed since the early nineties \citep[][]{dekool+ritter93-1}. 
The most important and, at the same time, least understood phase of compact binary evolution is CE evolution. 
The outcome of the CE phase is generally approximated by equating the binding energy of the envelope and the change in orbital energy scaled with an efficiency $ \alpha_{\mathrm{CE}}$, i.e., 
\begin{equation}\label{eq:alpha}
E_\mathrm{bind} = \alpha_{\mathrm{CE}}\Delta E_\mathrm{orb}.
\end{equation}
The most basic assumption is to approximate the binding energy only by the gravitational energy of the envelope:
\begin{equation}\label{eq:Egr}
E_\mathrm{bind} = E_\mathrm{grav} = −\frac{G M_\mathrm{1} M_\mathrm{1,e}}{\lambda R_\mathrm{1}},
\end{equation}
where $M_\mathrm{1}$, $M_\mathrm{1,e}$, and $R_\mathrm{1}$ are the total mass, envelope mass, and radius of the primary star, and $\lambda$ is a binding energy parameter that depends on the structure 
of the primary star. Previous simulations of PCEBs \citep{dekool+ritter93-1,willems+kolb04-1,politano+weiler06-1,politano+weiler07-1} have been performed using different values of the CE efficiency 
$\alpha_{\mathrm{CE}}$ but assuming $\lambda = 0.5$ or $\lambda = 1.0$, or assuming different fixed values for $\alpha_{\mathrm{CE}}\lambda$ \citep{toonen+nelemans13-1}. However, keeping $\lambda$ 
constant is not a very realistic assumption for all types of possible primaries, as pointed out by \citet{dewi+tauris00-1} and \citet{podsiadlowskietal03-1}. Very loosely bound envelopes in more evolved 
stars, e.g. if the primary is close to the tip of the AGB, can reach much higher values of $\lambda$. This is especially true if other sources of energy of the envelope, 
the most important being recombination energy,
support the ejection process. If a fraction $\alpha_{\mathrm{rec}}$ of the recombination energy of the envelope contributes to the ejection process, the binding energy equation becomes 
\begin{equation}\label{eq:Eball}
E_\mathrm{bind}=\int_{M_\mathrm{1,c}}^{M_\mathrm{1}}-\frac{G m}{r(m)}dm + \alpha_{\mathrm{rec}}\int_{M_\mathrm{1,c}}^{M_\mathrm{1}}U_\mathrm{rec}(m)
\end{equation}
where $M_\mathrm{1,c}$ is the core mass of the primary and $r(m)$  the radius that encloses the mass $m$. The effects of the additional energy source $U_{\mathrm{rec}}$ can 
be included in the $\lambda$ parameter by equating Eqs.\,(\ref{eq:Egr}) and\,(\ref{eq:Eball}). While it is clear that $\lambda$ is not constant, the contributions from other sources of energy, such as recombination, 
are very uncertain. On one hand, the existence of the long orbital-period PCEB IK\,Peg \citep{wonnacottetal93-1} might imply that there are missing 
terms in the energy equation, and the most promising candidate is indeed recombination energy available in the envelope \citep[see][ for a more detailed discussion]{webbink08-1}. 
On the other hand, it has been claimed that the opacity in the envelope is too low for an efficient use of recombination energy \citep{soker+harpaz03-1}. 

A first fairly rough attempt was made to investigate the impact of possible contributions of the recombination energy on the predictions of binary population models \citep{davisetal10-1}. 
However, the parameter space evaluated by these authors  was rather limited. First, they assumed $\alpha_{\mathrm{CE}} = 1.0$. Second, 
the values of $\lambda$ were obtained by interpolating the very sparse grid of \citet{dewi+tauris00-1}, which covered only eight primary masses and only the extreme cases 
of recombination energy contribution, i.e., $\alpha_{\mathrm{rec}}=0$ or $\alpha_{\mathrm{rec}}=1$. 

In this paper we simulate the population of detached WD+MS PCEBs with different values of the CE efficiency and with the inclusion of different fractions of recombination energy 
($\alpha_{\mathrm{rec}}$) in order to explore how these crucial parameters affect the properties of the predicted PCEB population.

\section{The simulations}\label{sec:sim}

We generate an initial MS+MS binary population of $10^7$ systems. The primary masses are distributed according to the initial mass function (IMF) of \citet{kroupaetal93-1}:
\begin{equation}
f(M_\mathrm{1}) = \left\{\begin{array}{l l}
  0 & \quad \mbox{$M_\mathrm{1}/\Msun<0.1,$}\\
  0.29056M_\mathrm{1}^{-1.3} & \quad \mbox{$0.1\leq{M_\mathrm{1}/\Msun}<0.5,$} \\
  0.15571M_\mathrm{1}^{-2.2} & \quad \mbox{$0.5\leq{M_\mathrm{1}/\Msun}<1.0,$} \\
  0.15571M_\mathrm{1}^{-2.7} & \quad \mbox{$1.0\leq{M_\mathrm{1}/\Msun}.$} \\
  \end{array}
  \right.
\label{M1dist}
\label{eq:IMF}
\end{equation}
For the mass of the secondary star we assume a flat initial-mass-ratio distribution (IMRD), i.e., $n(q)$ = constant, where $q = M_\mathrm{2}/M_\mathrm{1}$.
The initial orbital separation $a_\mathrm{i}$ follows the distribution
\begin{equation}
h(a_\mathrm{i}) = \left\{\begin{array}{l l}
  0 & \quad \mbox{$a_{\mathrm{i}}/\Rsun<3$ or $a_{\mathrm{i}}/\Rsun>10^{4},$}\\
  0.078636a_{\mathrm{i}}^{-1} & \quad \mbox{$3\leq a_{\mathrm{i}}/\Rsun \leq{10^4}$}\\ 
  \end{array}
  \right.
\label{adist}
\end{equation}
\citep{davisetal08-1}\footnote{\citet{davisetal08-1} give an upper limit of $10^{6}\Rsun$ for the distribution of initial separations. 
We cut the distribution at $10^{4}\Rsun$ because in systems with larger initial separations, the primary will never fill the Roche lobe.}.
We assumed solar metallicity for all the systems.
Finally we assign a ``born time'' ($t_{\mathrm{born}}$) to all the systems, corresponding to the age of the Galaxy when the system was born, assuming a 
constant star formation rate between $0$ and the age of the Galaxy ($t_{\mathrm{Gal}} \sim\,13.5$\,Gyr).

We use the latest version of the binary-star evolution (BSE) code from \citet{hurleyetal02-1}, updated as in \citet{zorotovic+schreiber13-1},  to evolve all 
the systems during $t_{\mathrm{evol}} = t_{\mathrm{Gal}} - t_{\mathrm{born}}$, in order to obtain the {\em{current}} orbital and stellar parameters. 
Disrupted magnetic braking is assumed. As discussed in detail in \citet{zorotovicetal10-1}, the latest version of the BSE code allows one to compute the binding 
energy of the envelope, including not only the gravitational energy but also a fraction $\alpha_{\mathrm{rec}}$ of the recombination energy of the envelope. The two 
free parameters in our simulations are then the CE efficiency ($\alpha_{\mathrm{CE}}$) and the fraction of recombination energy that is used to expel the envelope 
($\alpha_{\mathrm{rec}}$). 

\begin{table}
\caption{\label{tab:mod} Different models analyzed in this work.}
\begin{center}
\begin{tabular}{lccc}
\hline\hline
Model  & $\alpha_{\mathrm{CE}}$ & $\alpha_{\mathrm{rec}}$\\
\hline
a & 0.25 & 0.00\\
b & 0.25 & 0.02\\
c & 0.25 & 0.25\\
d & 0.50 & 0.00\\
e & 0.50 & 0.02\\
f & 0.50 & 0.25\\
g & 1.00 & 0.00\\
h & 1.00 & 0.02\\
i & 1.00 & 0.25\\
\hline
\noalign{\smallskip}
\end{tabular}
\end{center}
\end{table}

We assume that the fraction of recombination energy that contributes to the envelope ejection process cannot exceed the efficiency of using the orbital 
energy of the binary. This is reasonable because the recombination energy is probably radiated away much more easily. Table\,\ref{tab:mod} shows the combination of 
the two efficiency parameters for the nine different models we studied in this work.

Our simulated PCEB sample contains all the WD+MS binaries that went through a CE phase but did not yet reach the second phase of mass transfer, which 
would probably make them cataclysmic variables.

\section{Results} 

In what follows we describe and explain the characteristics of the predicted parameter distributions for the nine models listed in Table\,\ref{tab:mod}. 

\subsection{Number of PCEBs}

Table\,\ref{tab:res} lists the total number of detached WD+MS PCEBs predicted by each model\footnote{The total number of systems obtained for each model
is not a prediction of what should be expected observationally, and should not be used to estimate space densities. It is only listed to show how increasing 
both efficiencies allows more systems to survive the CE phase.}, 
as well as the fractions of systems containing He-core and C/O-core WDs. The total number of systems increases noticeably with the value of $\alpha_{\mathrm{CE}}$ and also slightly with 
the value of $\alpha_{\mathrm{rec}}$. This is easy to understand: a higher value of the CE efficiency implies a more efficient use of orbital energy and thus a 
smaller reduction of the binary separation, which allows more systems to survive. In addition, systems that survive with a low  CE efficiency emerge from the CE 
phase at longer periods when we increase the efficiency, and therefore stay longer as detached PCEBs. The same occurs if an increasing fraction of recombination energy 
is assumed to contribute. However, $\alpha_{\mathrm{rec}}$ does not affect all the systems in the same way, because the relative contribution of recombination energy 
depends on the mass and evolutionary state of the primary. For example, for less evolved primaries on the FGB, the contribution of recombination energy to the binding 
energy remains small compared to the contribution of gravitational energy even for high values of $\alpha_{\mathrm{rec}}$, because the envelope is not as extended 
as in the AGB and is still tightly bound to the core.

\begin{table}
\caption{\label{tab:res} Results for $n(q)$ = constant.}
\begin{center}
\begin{tabular}{lccccc}
\hline\hline
Model & $\alpha_{\mathrm{CE}}$ & $\alpha_{\mathrm{rec}}$ & $N_{\mathrm{sys}}$ & He (\%) & C/O (\%)\\
\hline
a & 0.25 & 0.00 & 33\,917 & 44.6 & 55.4 \\
b & 0.25 & 0.02 & 36\,098 & 42.8 & 57.2 \\
c & 0.25 & 0.25 & 45\,279 & 41.3 & 58.7 \\
d & 0.50 & 0.00 & 60\,444 & 51.3 & 48.7 \\
e & 0.50 & 0.02 & 61\,745 & 50.6 & 49.4 \\
f & 0.50 & 0.25 & 68\,215 & 49.7 & 50.3 \\
g & 1.00 & 0.00 & 88\,039 & 56.1 & 43.9 \\
h & 1.00 & 0.02 & 88\,886 & 55.7 & 44.3 \\
i & 1.00 & 0.25 & 92\,726 & 55.2 & 44.8 \\
\hline
\end{tabular}
\end{center}
\tiny Total number of detached PCEBs obtained with each model and percentage of systems with He WDs and C/O WDs. From the $10^7$ initial MS+MS binaries simulated with this distribution, $\sim40.7$\% entered a CE phase.
\end{table}

\begin{figure*}
\centering
\includegraphics[angle=270,width=0.59\textwidth]{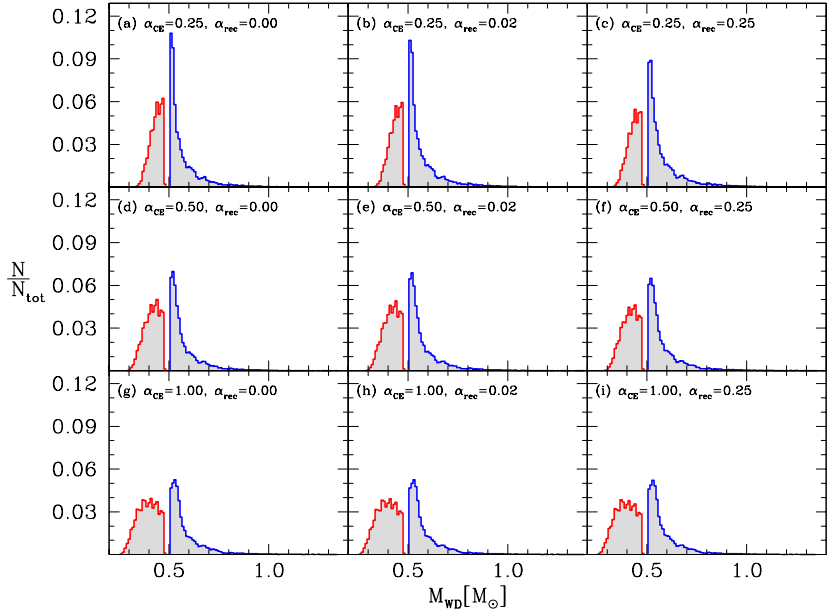}
\caption{WD mass distribution for the different models. Gray shaded histograms represent the entire distribution, while the color histograms are for systems with He WDs 
(red) and with C/O WDs (blue). 
}
\label{fig:Mwd}
\end{figure*}

\begin{figure*}
\centering
\includegraphics[angle=270,width=0.59\textwidth]{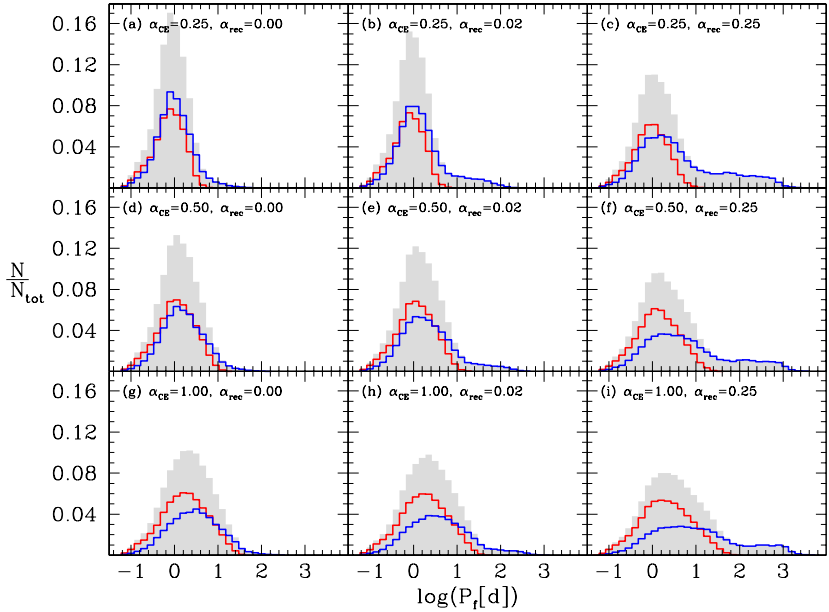}
\caption{Orbital period distribution for the different models described in Table\,\ref{tab:mod}. Colors are the same as in Fig.\,\ref{fig:Mwd}. 
}
\label{fig:Pf}
\end{figure*}

\begin{figure*}
\centering
\includegraphics[angle=270,width=0.59\textwidth]{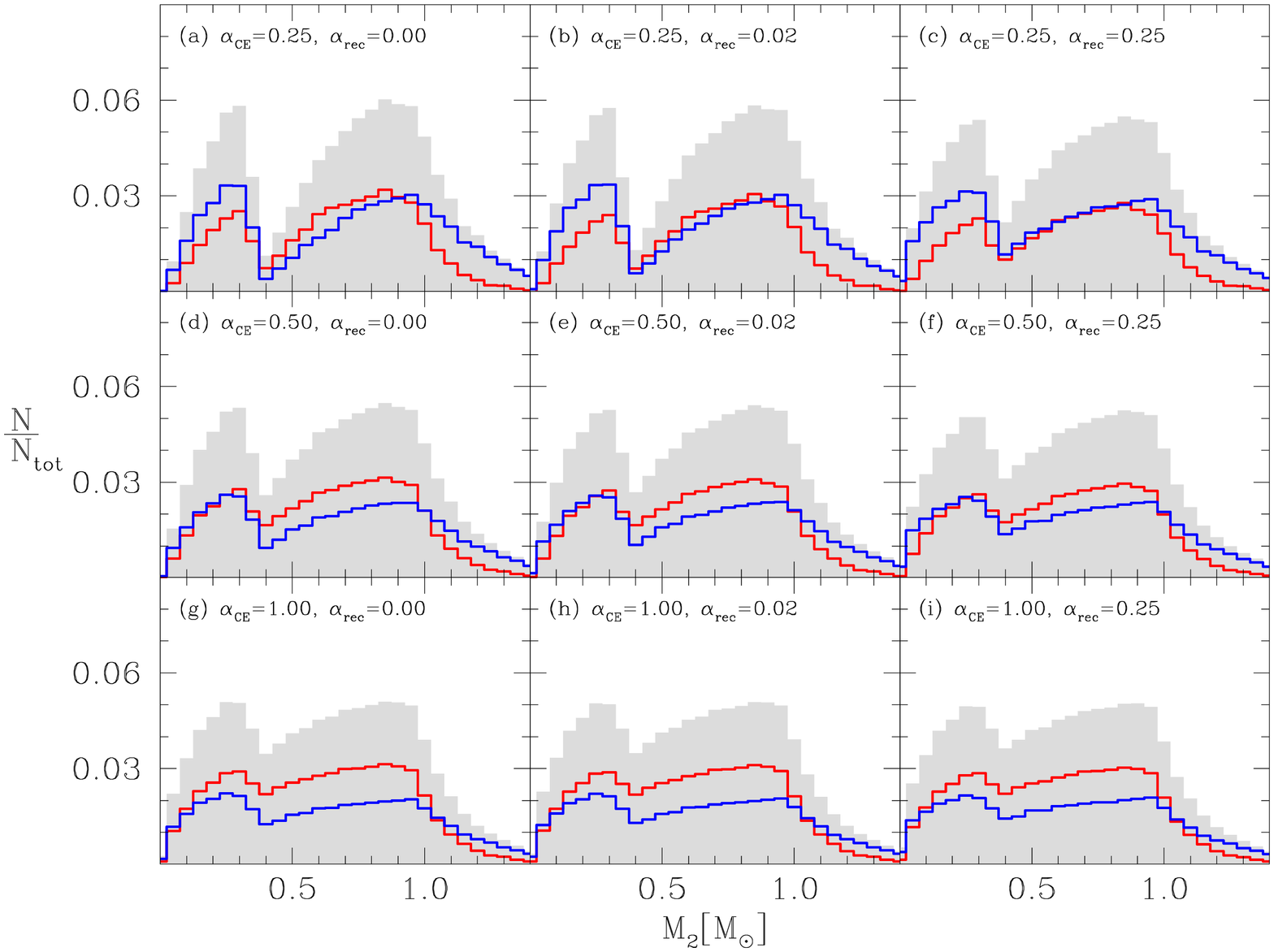}
\caption{Secondary mass distribution for the different models. Colors are the same as in Fig.\,\ref{fig:Mwd}. 
}
\label{fig:M2}
\end{figure*}

\subsection{The WD mass distribution}

Figure\,\ref{fig:Mwd} and Table\,\ref{tab:res} show the WD mass distribution for all the models. The gap separating systems with He WDs from those containing C/O WDs is caused by the stellar 
radius at the tip of the FGB being larger than at the beginning of the AGB, while the core mass still increases from $\sim0.48$ to $0.51\Msun$. In this range of 
core masses, the primary star cannot fill its Roche lobe because it would have done so before on the FGB. 

Figure\,\ref{fig:Mwd} clearly shows that the relative number of systems with He WDs increases and that the distribution extends towards lower mass systems for higher 
values of $\alpha_{\mathrm{CE}}$. Less evolved systems, like those in which the primary star is a (low-mass) He WD, are initially closer and therefore a lower value 
of $\alpha_{\mathrm{CE}}$ implies an increased merger rate for these systems, while the progenitors of systems with C/O WDs are initially more separated and can survive 
the CE evolution even if more orbital energy is required to expel the envelope (small $\alpha_{\mathrm{CE}}$). Therefore, the shape of the WD mass distribution for 
systems containing high-mass C/O WDs is almost unaffected by the value of the CE efficiency. For a fixed value of $\alpha_{\mathrm{CE}}$, on the other hand, 
the percentage of systems containing a He WD remains nearly constant (with a very slight decrease) for different values of $\alpha_{\mathrm{rec}}$. This is because 
the recombination energy becomes more important than the gravitational energy
only for very advanced evolutionary stages, especially later on the AGB. For those systems, the initial separation is generally 
large enough to avoid a merger even without including this additional energy.  

\subsection{The orbital period distribution}

The orbital period distributions predicted by our nine models are shown in Fig.\,\ref{fig:Pf}. The orbital periods for systems containing C/O WDs are on average 
longer than those of systems containing He WDs in all the models. The peak of the period distributions for the entire sample shifts toward longer periods 
if $\alpha_{\mathrm{CE}}$ is increased. Also, by increasing the value of $\alpha_{\mathrm{CE}}$, the distribution becomes slightly wider. This 
is because greater CE efficiency implies a smaller reduction of the orbital period, moving the distributions toward longer orbital periods but also adding new 
systems with short periods that mainly contain He WDs. These systems merge for low values of $\alpha_{\mathrm{CE}}$ but can survive the CE phase if the orbital 
energy is used efficiently. The effect of increasing the fraction of recombination energy mostly affects systems with longer periods and C/O WDs that descend from 
evolved primaries where the contribution of the recombination energy of the envelope becomes important. A tail toward longer orbital periods appears in the distribution 
for systems with C/O WDs with increasing $\alpha_{\mathrm{rec}}$, while the shape of the distribution for systems with He WDs remains nearly constant for a fixed 
value of $\alpha_{\mathrm{CE}}$. Almost all the systems with periods longer than about ten days can only be obtained when a fraction of the recombination energy is taken into account. 

\subsection{The secondary mass distribution}

In Fig.\,\ref{fig:M2} we show the distributions obtained for the secondary masses. The relative number of systems increases with increasing secondary mass, with a steep 
decline at $M_2\sim\,0.35$\Msun. This corresponds to the boundary for fully convective secondaries where, according to the disrupted magnetic braking theory, 
angular momentum loss due to magnetic braking becomes inefficient. A PCEB evolves toward shorter orbital periods because of orbital angular momentum loss through 
gravitational radiation and the much stronger magnetic wind braking. Below $M_2\sim\,0.35$\Msun\, PCEBs get closer only thanks to gravitational radiation, which is 
much less efficient than magnetic braking, causing these systems to spend more time as detached PCEBs before the secondary fills its Roche lobe and becomes a 
cataclysmic variable, and therefore increasing the relative number of systems with low-mass secondaries. This behavior has already been predicted by \citet{politano+weiler07-1} 
and observationally confirmed by \citet{schreiberetal10-1}. The effect of increasing $\alpha_{\mathrm{CE}}$ is that this decline becomes less apparent. This is because the 
distributions are normalized for each model, and as already mentioned, increasing $\alpha_{\mathrm{CE}}$ rapidly increases the number of systems obtained 
(see Table\,\ref{tab:res}) and moves the orbital period distribution toward longer periods. More systems therefore stay as detached PCEBs for very long periods of time, 
up to several Hubble times, even when magnetic braking is efficient ($M_2\gappr\,0.35$\Msun\,). The effect of increasing $\alpha_{\mathrm{rec}}$ is similar but much less 
pronounced because recombination energy mainly affects systems with more evolved primaries. The drop of systems toward masses higher than $\sim\,1$\Msun\, is the imprint 
of the IMF for the primary, because $M_2$ is related to $M_1$ through the IMRD. 

\subsection{Relating the final parameters}\label{sec:relate}

\begin{figure*}
\centering
\includegraphics[angle=270,width=0.59\textwidth]{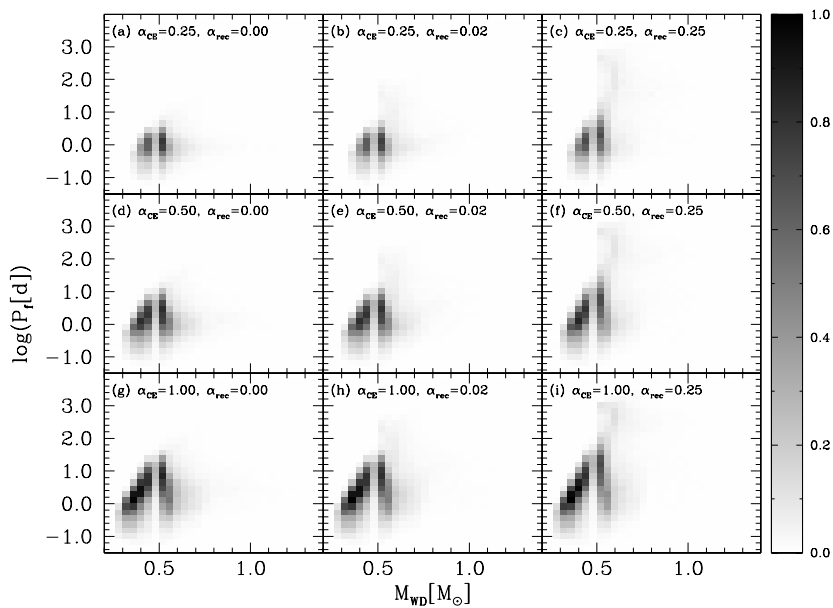}
\caption{Relation between WD mass and orbital period. The intensity of the gray scale represents the density of objects in each bin, on a linear scale, and normalized 
to one for the bin that contains most systems. 
} 
\label{fig:PMwd}
\end{figure*}

\begin{figure*}
\centering
\includegraphics[angle=270,width=0.59\textwidth]{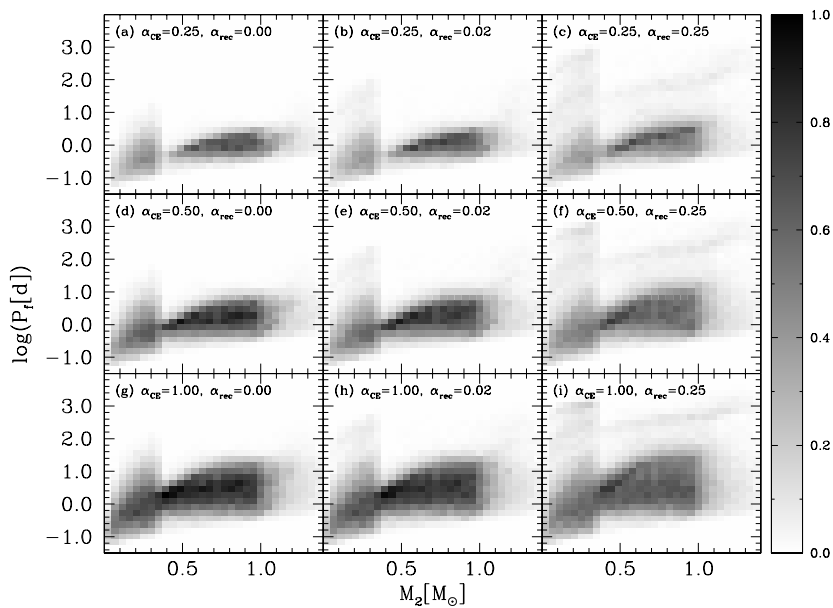}
\caption{Relation between secondary mass and orbital period. The intensity of the gray scale means the same as in Fig.\,\ref{fig:PMwd}. 
}
\label{fig:PM2}
\end{figure*}

\begin{figure*} 
\centering
\includegraphics[angle=270,width=0.59\textwidth]{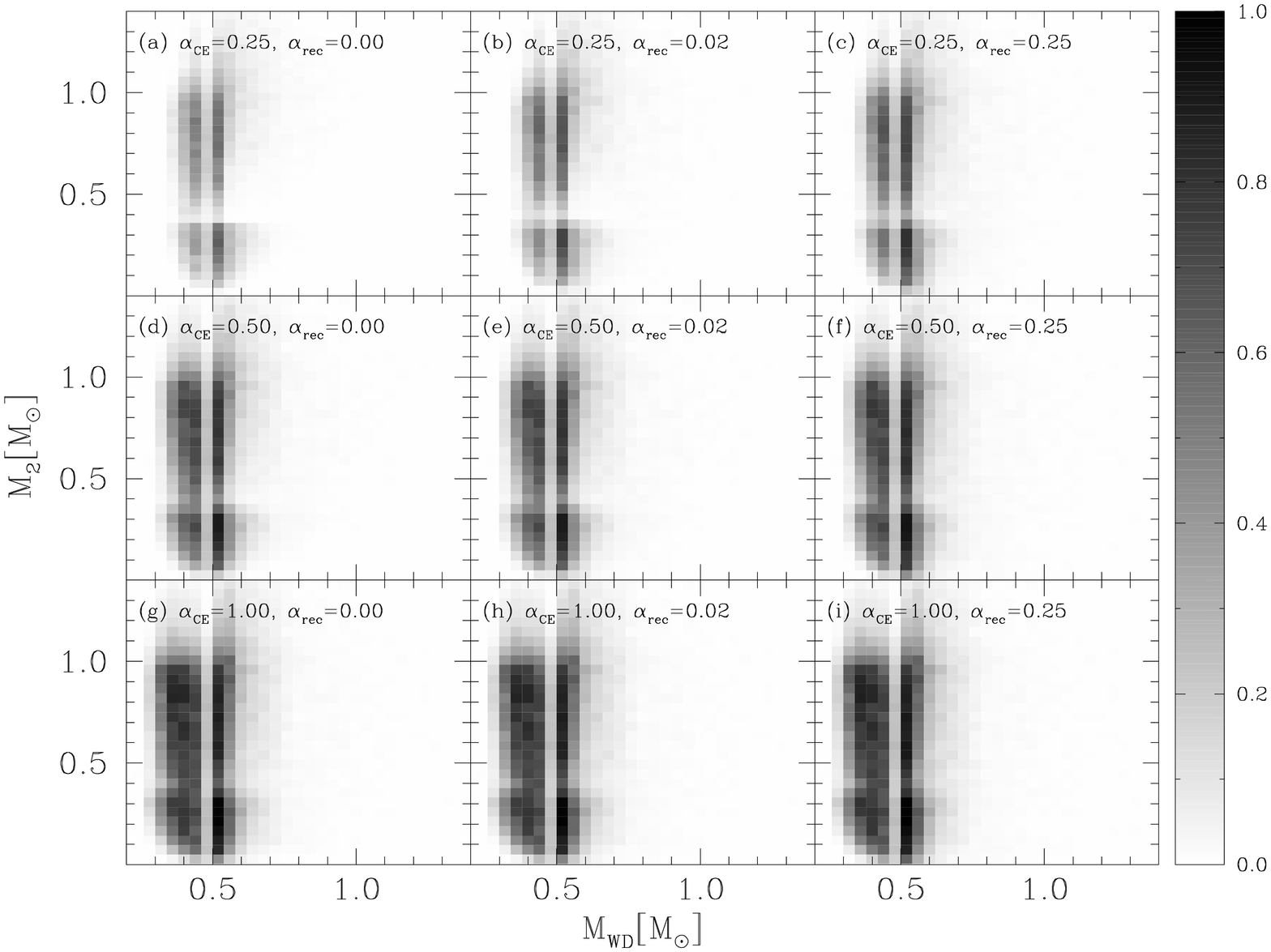}
\caption{Relation between WD and secondary mass. The intensity of the gray scale means the same as in Fig.\,\ref{fig:PMwd}. 
}
\label{fig:MwdM2}
\end{figure*}

In addition to inspecting distributions of a single parameter, it is instructive to investigate possible relations between the orbital and stellar parameters. 
Figure\,\ref{fig:PMwd} shows the relation between the WD mass and the orbital period. The gap separating systems with He WDs from systems with C/O WDs is evident. 

Among the systems with He WDs, there is a correlation between the orbital period and the WD mass, a trend that becomes more apparent by increasing
$\alpha_{\mathrm{CE}}$ as systems with lower mass He WDs survive. In contrast, no clear
trend can be identified
for systems with C/O WDs. This difference agrees with the observations \citep{zorotovicetal11-2} and can be understood as follows. 
The C/O WDs in PCEBs descend from a wider range of progenitor masses and initial separations \citep[see][ their figure 2]{zorotovic+schreiber13-1},
which also results in a wider range of masses for the companion.
This translates into a wider range of initial energies (orbital and binding) and values of the binding energy parameter $\lambda$ (especially if the effects of recombination energy are included). 
This wider range of initial conditions naturally transfers into a wider range of final orbital periods for systems containing C/O WDs with similar masses.
In particular, the strong impact of potential contributions of recombination energy on the final periods of PCEBs containing C/O WDs is clearly visible in Fig.\,\ref{fig:PMwd}.
Increasing the fraction of recombination energy that is used to expel the envelope mainly affects those systems with more massive C/O WDs, 
where the value of $\lambda$ can become extremely high, moving them toward longer periods. Therefore, as pointed out previously by \citet{rebassa-mansergasetal12-1}, 
clear observational constraints on the role of the recombination energy could be derived eventually if the orbital periods of a large and 
representative sample of PCEBs containing high-mass WDs could be measured. 

In Fig.\,\ref{fig:PM2} we show the relation between the mass of the secondary star and the orbital period. There is a tendency to predict longer periods for 
systems with more massive secondaries in agreement with the observational analysis of \citet{zorotovicetal11-2}. The reason for this is 
twofold. First, for a given primary mass and orbital period, more initial orbital energy is available for systems with more massive secondaries, and therefore the fraction of this 
energy that is needed to unbind the envelope is smaller, leading to longer orbital periods. Second, for a given WD mass, the minimum period at which a
PCEB remains detached decreases with secondary mass. Since lower mass secondaries have smaller radii, they can remain within their Roche lobes at smaller
separations (shorter orbital periods).  

The previously mentioned paucity of systems with $M_2\sim0.35-0.5$\Msun\,is also evident in Fig.\,\ref{fig:PM2}.
Owing to the assumption of disrupted magnetic braking in our simulations, PCEBs with masses exceeding $\sim0.35$\Msun\, become closer not only because of gravitational 
radiation but also due to magnetic braking, which is supposed to be much more efficient. This causes much shorter evolutionary time scales from the 
CE to the CV phase. This explains the reduction of systems with secondary masses exceeding the fully convective boundary located at $0.35$\Msun. 
In the range $M_2\sim0.35-0.5$\Msun\, almost all systems with long ($\log\Porb[d]>0.5$) and short orbital periods ($\log\Porb[d]<-0.5$) disappeared. 
At $\log\Porb[d]\sim0$, a significant number of systems with $M_2\sim0.35-0.5$\Msun\  remain despite the efficient angular momentum loss due
to magnetic braking because of the very large number of PCEBs formed with these parameters (for a flat IMRD as assumed here). 

Figure\,\ref{fig:PM2} also nicely shows that increasing the values of
$\alpha_{\mathrm{CE}}$ or $\alpha_{\mathrm{rec}}$ reduces the paucity of
systems with 
$\sim0.35-0.5$\Msun\, 
secondary stars (caused by assuming disrupted magnetic braking). 
This is because the PCEBs emerge from CE evolution with longer orbital periods and remain longer as detached systems,
which increases the total number of PCEBs, even if the mass of the secondary star implies magnetic braking to be efficient. 
It can also be seen that the increase in long-period systems due to higher values of $\alpha_{\mathrm{rec}}$ is independent of secondary mass. 

Finally, Fig.\,\ref{fig:MwdM2} shows the relation between the masses of the WD and the secondary star. The three previously mentioned features can be identified as well, i.e. the increase 
in the total number of systems with increasing $\alpha_{\mathrm{CE}}$, the increase of systems with low-mass He WDs for increasing $\alpha_{\mathrm{CE}}$, and the less apparent 
decline in the number of systems with masses  $\sim0.35-0.5$\Msun\, as $\alpha_{\mathrm{CE}}$ or $\alpha_{\mathrm{rec}}$ are increased. In agreement with the observational 
findings from \citet{zorotovicetal11-2}, there seems to be no relation between the two stellar masses.

\subsection{The initial-mass-ratio distribution}

To test whether the IMRD has any effect on the period and mass
distributions, we decided to repeat our full set of simulations assuming 
different IMRDs. 
And assuming two extreme cases, i.e. $n(q)\propto q$, in addition to $n(q)\propto q^{-1}$, we obtained the following results.

The total number of detached WD+MS PCEBs predicted by each model and the fractions of systems containing He and C/O WDs are shown in
Tables\,\ref{tab:res2} and\,\ref{tab:res3} for the additional IMRDs. The fraction of systems entering a CE phase is virtually independent
of the IMRD, because the assumed initial mass function for the primary and the distribution of initial separations are identical in all simulations 
and dominate the weak dependence of the Roche-lobe radius of the primary on the secondary mass.
For the two new IMRDs, the total number of systems increases markedly with 
$\alpha_{\mathrm{CE}}$ and also somewhat with $\alpha_{\mathrm{rec}}$, as in the case of a flat distribution (see Table\,\ref{tab:res}).
The simulations that assume an IMRD inversely proportional to $q$ generate more WD+MS PCEBs than in the case of a flat distribution, while
simulations assuming $n(q)\propto q$ generate less systems. This can be explained as a combination of two effects. Assuming $n(q)\propto q^{-1}$ favors 
the formation of systems with low-mass secondary stars, which take longer to evolve and therefore remain longer as MS stars. On the other hand, 
more massive secondaries may have enough time to evolve, and then the system will no longer be a WD+MS PCEB. Also, if
the mass of the secondary is smaller than $\sim\,0.35\,\Msun$ the system remains detached after the CE phase for longer,
because magnetic braking is not acting (or at least not efficiently acting) and angular momentum loss is driven mainly due to gravitational radiation.

For both distributions, the fraction of systems with He or C/O WDs behaves in the same way as for a flat IMRD; i.e., the fraction of systems with He WDs 
increases notably by increasing $\alpha_{\mathrm{CE}}$ and slightly decreases by increasing $\alpha_{\mathrm{rec}}$.
The fraction of systems with He WDs is greater for the distribution favoring more massive secondary stars ($n(q)\propto q$).
This is for several reasons. First and most important, systems with more massive secondaries have a 
higher initial orbital energy, before the CE phase, and therefore have more energy available to unbind the envelope. Systems where the 
envelope relatively tightly bound, such as the progenitors of He WDs, can survive the CE phase
more easily if they have a massive companion. 
Second, systems with more massive secondaries emerge from the CE phase with a longer orbital periods and therefore 
remain detached PCEBs for longer. This increases the fraction of systems with He WDs because these are the ones that end up closer after 
the CE phase and start a second phase of mass transfer faster earlier. Finally, there is also a tendency to produce slightly less massive WDs 
in systems with more massive secondaries because, for a given primary mass, the Roche lobe of the primary is smaller when 
the secondary star is more massive.

\begin{table}
\caption{\label{tab:res2} Results for $n(q)\propto q$.}
\begin{center}
\begin{tabular}{lccccc}
\hline\hline
Model & $\alpha_{\mathrm{CE}}$ & $\alpha_{\mathrm{rec}}$ & $N_{\mathrm{sys}}$ & He (\%) & C/O (\%)\\
\hline
a & 0.25 & 0.00 & 30\,195 & 51.6 & 48.4 \\
b & 0.25 & 0.02 & 31\,188 & 50.9 & 49.1 \\
c & 0.25 & 0.25 & 36\,712 & 50.4 & 49.6 \\
d & 0.50 & 0.00 & 50\,090 & 59.2 & 40.8 \\
e & 0.50 & 0.02 & 50\,087 & 58.7 & 41.3 \\
f & 0.50 & 0.25 & 54\,589 & 58.6 & 41.4 \\
g & 1.00 & 0.00 & 70\,035 & 64.2 & 35.8 \\
h & 1.00 & 0.02 & 70\,490 & 63.9 & 36.1 \\
i & 1.00 & 0.25 & 72\,474 & 63.6 & 36.4 \\
\hline
\end{tabular}
\end{center}
\tiny Same as in Table\,\ref{tab:res} but for the IMRD proportional to the mass ratio. From the $10^7$  initial MS+MS binaries simulated with this distribution, $\sim40.4$\% entered a CE phase.
\end{table}

\begin{table}
\caption{\label{tab:res3} Results for $n(q)\propto q^{-1}$.}
\begin{center}
\begin{tabular}{lccccc}
\hline\hline
Model & $\alpha_{\mathrm{CE}}$ & $\alpha_{\mathrm{rec}}$ & $N_{\mathrm{sys}}$ & He (\%) & C/O (\%)\\
\hline
a & 0.25 & 0.00 & 38\,680 & 39.0 & 61.0 \\
b & 0.25 & 0.02 & 42\,515 & 36.0 & 64.0 \\
c & 0.25 & 0.25 & 55\,572 & 34.4 & 65.6 \\
d & 0.50 & 0.00 & 72\,625 & 45.5 & 54.5 \\
e & 0.50 & 0.02 & 74\,989 & 44.3 & 55.7 \\
f & 0.50 & 0.25 & 83\,632 & 43.1 & 56.9 \\
g & 1.00 & 0.00 & 109\,711 & 49.6 & 50.4 \\
h & 1.00 & 0.02 & 110\,629 & 49.3 & 50.7 \\
i & 1.00 & 0.25 & 115\,953 & 49.0 & 51.0 \\
\hline
\end{tabular}
\end{center}
\tiny Same as in Table\,\ref{tab:res} but for the IMRD inversely proportional to the mass ratio. From the $10^7$ initial MS+MS binaries simulated with this distribution, $\sim41.0$\% entered a CE phase.
\end{table}

The WD mass distribution is almost unaffected by the assumption of a different IMRD. The shape of the two distributions for systems containing
He and C/O WDs remains almost identical with the only variation being their relative contributions to the whole population.
This was expected because, as mentioned in Sect.\,\ref{sec:relate}, both masses do not appear to be related (see also Fig.\,\ref{fig:MwdM2}). 

\begin{figure*}
\centering
\includegraphics[angle=270,width=0.59\textwidth]{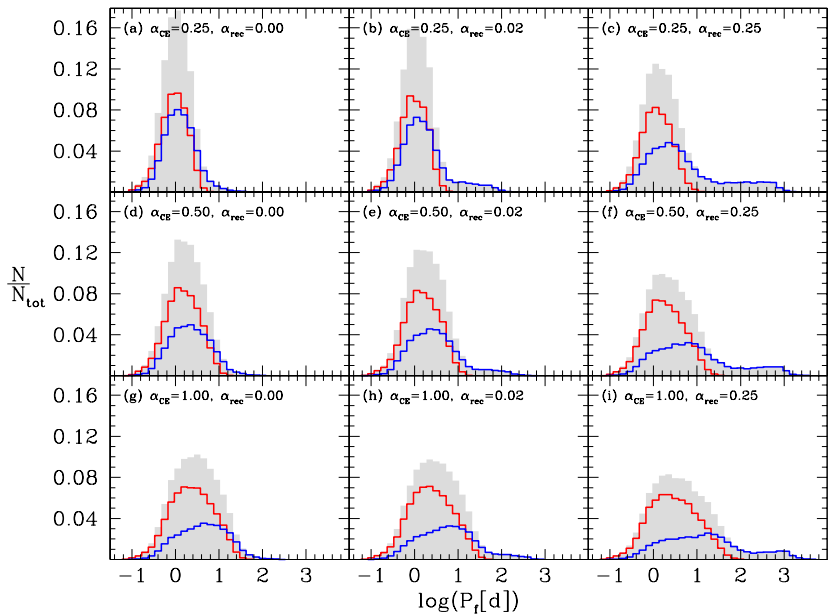}
\caption{Same as in Fig.\,\ref{fig:Pf} but for $n(q)\propto q$. 
}
\label{fig:Pf_prop}
\end{figure*}

\begin{figure*}
\centering
\includegraphics[angle=270,width=0.59\textwidth]{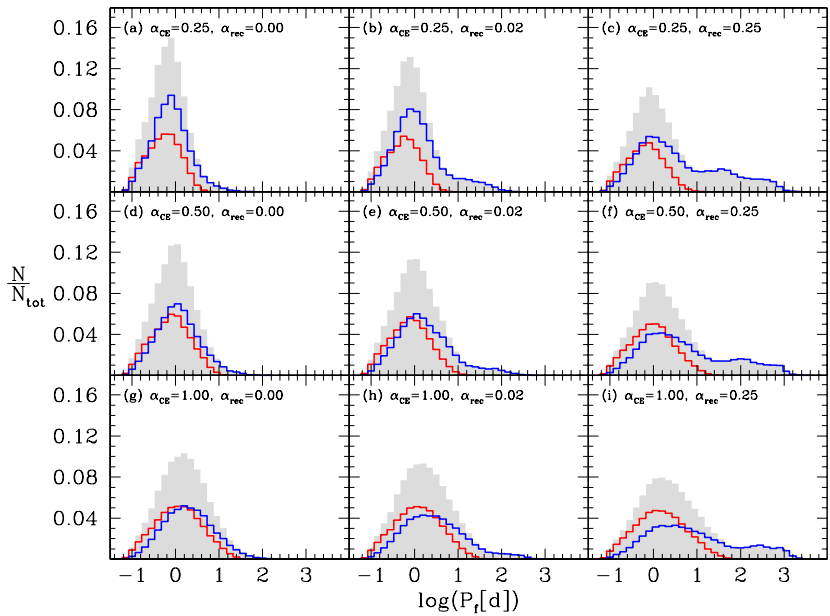}
\caption{Same as in Fig.\,\ref{fig:Pf} but for $n(q)\propto q^{-1}$. 
}
\label{fig:Pf_inv}
\end{figure*}

The period distributions are shown in Figs.\,\ref{fig:Pf_prop} and\,\ref{fig:Pf_inv} for the IMRD proportional to the mass ratio
and for the one in which the secondary mass depends inversely on the mass ratio, respectively.
The shape of the distributions does not change dramatically by using a different IMRD. However, the entire distributions move slightly
toward longer (shorter) orbital periods when we favor the formation of systems with more (less) massive secondaries, respectively. This is because, as we show in Fig.\,\ref{fig:PM2},
there is a relation between the mass of the secondary and the orbital period; i.e., systems with more massive secondaries tend to have longer
periods.

\begin{figure*}
\centering
\includegraphics[angle=270,width=0.59\textwidth]{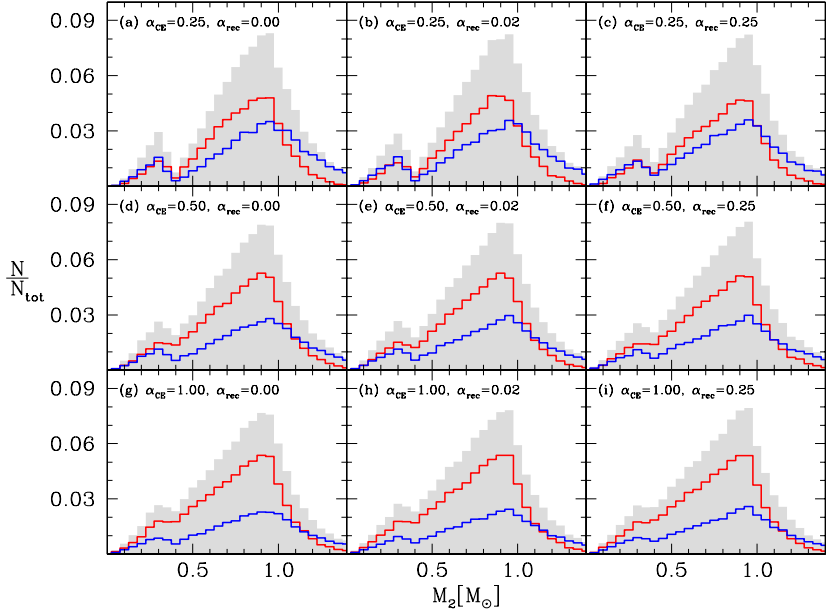}
\caption{Same as in Fig.\,\ref{fig:M2} but for $n(q)\propto q$. 
}
\label{fig:M2_prop}
\end{figure*}

\begin{figure*}
\centering
\includegraphics[angle=270,width=0.59\textwidth]{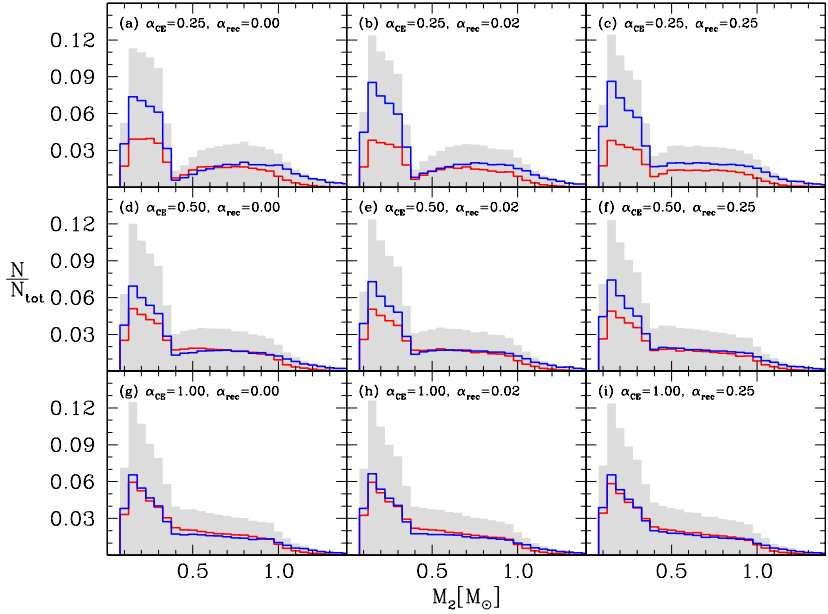}
\caption{Same as in Fig.\,\ref{fig:M2} but for $n(q)\propto q^{-1}$. 
}
\label{fig:M2_inv}
\end{figure*}

Figures\,\ref{fig:M2_prop} and\,\ref{fig:M2_inv} show the distributions of secondary masses for the cases in which
$n(q)\propto q$ and $n(q)\propto q^{-1}$, respectively.  
While for the case of a flat IMRD the two peaks in this distributions have approximately the same height (see Fig.\,\ref{fig:M2}), 
it is evident from these two figures that we are favoring the formation of systems with high- and low-mass secondaries, respectively.
As in the case of the flat IMRD, the steep decline at the boundary for fully convective secondaries is more pronounced for low 
values of $\alpha_{\mathrm{CE}}$, and it becomes almost indistinguishable when we increase the value of $\alpha_{\mathrm{CE}}$
for the models in which we assume $n(q)\propto q$.
If one could have a homogeneous and unbiased sample of WD+MS PCEBs, covering the whole range of masses for the companion star, the distribution
of secondary masses would be very useful for deriving constraints on the IMRD. 

\section{Discussion}

We have performed detailed binary population simulations of detached WD+MS binaries that evolved through CE evolution for different CE efficiencies $\alpha_{\mathrm{CE}}$. 
For the first time we have done a systematic and comprehensive study of the effects of the recombination energy parametrized with $\alpha_{\mathrm{rec}}$. In what follows we discuss the 
predictions of our model in the context of previous model calculations. 

The first detailed simulations of WD+MS PCEBs were performed by \citet{dekool+ritter93-1}, and several of their predictions are still valid; for example, the decrease in the relative 
number of PCEBs with He WDs for lower values of the CE efficiencies goes back to this early work. However, \citet{dekool+ritter93-1} just used a relatively small set of 
different parameters and assumed a constant binding energy parameter $\lambda=0.5,$ which is not always a realistic assumption \citep{dewi+tauris00-1}. More than a decade later, 
\citet{willems+kolb04-1} updated and extended the early work of \citet{dekool+ritter93-1} by covering a larger parameter space and using more recent fits to stellar evolutionary 
sequences \citep{hurleyetal00-1}. The predictions presented in these early papers are, however, difficult to compare with the observations because only current zero-age PCEB 
distributions were calculated; i.e., the evolution of PCEBs toward shorter orbital periods was not taken into account. 

\citet{politano+weiler07-1} were the first to present a predicted present-day population of PCEBs (their figures 2-5) to investigate the impact of assuming very low values 
of the CE efficiency (i.e., $\alpha_{\mathrm{CE}}<0.2$) and a dependence of $\alpha_{\mathrm{CE}}$ on the mass of the secondary star. Our simulations agree with their predictions
with respect to the reduced number of He WD primaries for low CE efficiencies and to the existence of less massive He WDs for higher values ​​of $\alpha_{\mathrm{CE}}$ 
(bottom panels in their Fig.\,3 and our Fig.\,1) and with the more pronounced decrease at the fully convective boundary in the distribution of the secondary masses 
(top panels in their Fig.\,3 and our Fig.\,3). Later, \citet{davisetal10-1} performed comprehensive binary population simulations of PCEBs for 
the first time taking into account that the binding energy parameter is probably not a constant. They find that the predicted distributions agree reasonably well with the observed populations 
for a constant value of $\alpha_{\mathrm{CE}}$ but predict a tail of long orbital period systems that was not present in the observed sample available to them. 

Finally, in 
a very recent work, \citet{toonen+nelemans13-1} simulated the current population of PCEBs in the Galaxy taking observational biases specific to the Sloan
Digital Sky Survey (SDSS) into account. They find a better fit to the observations by using a low value of $\alpha_{\mathrm{CE}}$ (0.25), which is consistent with the results from 
\citet{zorotovicetal10-1}. However, the fraction of systems containing He WD primaries is too high in their simulations. They suggest that this can be solved by using a 
higher value of $\alpha_{\mathrm{CE}}$ when the CE phase begins during the AGB. However, this study also did not include the effects of recombination energy and adopted a constant 
value for $\alpha_{\mathrm{CE}}\lambda$, which as outlined above, is not always a good approximation because $\lambda$ depends on the properties of the star, 
in particular on its mass and radius \citep[see,
  e.g.,][]{dewi+tauris00-1}. Although a constant value might be a good
approximation for most systems, this becomes completely unrealistic
for systems where the primary filled the Roche 
lobe at a more advanced evolutionary stage, with a less tightly bound 
envelope.
We emphasize at this point that one therefore needs to be careful when 
drawing conclusions based on the assumption of $\alpha_{\mathrm{CE}}\lambda$ 
constant.

Here we extended the study of \citet{politano+weiler07-1}, \citet{davisetal10-1}, and \citet{toonen+nelemans13-1} by presenting the first systematic investigation that 
includes the contribution from recombination energy to the energy budget of CE evolution. 

\section{Conclusions}

We have performed binary population synthesis models of PCEBs that include the possible contribution of recombination energy during CE evolution. The main features that characterize 
the distributions of the orbital parameters for the different models can be summarized as follows:
\begin{itemize}
\item The orbital period distributions become slightly wider by increasing the value of $\alpha_{\mathrm{CE}}$. 
\item Including a fraction of the recombination energy mainly affects systems with the more massive C/O WDs by producing a tail in the period distribution toward longer orbital periods.
\item The fraction of systems with He WDs increases by increasing $\alpha_{\mathrm{CE}}$, and the distribution extends toward lower mass systems ($\lappr\,0.3\,\Msun$).
\item The distribution of secondary masses has a steep decline at $M_2\sim\,0.35\,\Msun$, as a consequence of assuming disrupted magnetic braking, which is more pronounced 
for low values of $\alpha_{\mathrm{rec}}$ and especially of $\alpha_{\mathrm{CE}}$. 
\item Systems with more massive secondaries tend to have longer periods after
  the CE phase in all models. 
\item The predicted distribution of secondary masses is very similar for different WD masses. The distribution changes with the IMRD; i.e., if initially high mass ratios are favored, all WDs have 
larger numbers of relatively massive companions. If instead low initial mass ratios dominate, all WDs (independent of their mass) are more frequently found to have low-mass companions.
\item The relation between the period and the mass of the secondary means that
  the period distribution moves slightly toward longer orbital periods when
  we assumed an IMRD that favors the formation of systems with massive companions.
\item The mass distribution of the secondaries is strongly affected by the choice of the IMRD.

\end{itemize}

Some of these features may be used in combination with a large observational sample to put constraints on the values of $\alpha_{\mathrm{CE}}$ and/or $\alpha_{\mathrm{rec}}$,
as well as on the IMRD. 
A detailed analysis of the selection effects that affect the sample of WD+MS PCEBs obtained from the SDSS, as well as a thorough comparison with the observed sample of these systems, 
was recently presented by \citet{camachoetal2014-1}. While the best agreement between observations and theory has been found for low values of $\alpha_{\mathrm{CE}}\sim0.25$, the observed sample
is still too small to derive robust constraints. This is mostly for three reasons. First, the spectroscopic SDSS survey allows one to identify only low-mass companions (spectral type M) 
to WDs. Second, the performed radial velocity survey somewhat favors the detection of short orbital period systems. Third, after taking the observational biases and selection
effects into account, a relatively small sample of observed systems remained. Once a large and homogeneous sample of PCEBs is known, we recommend the following diagnostics to constrain currently unknown parameters. 

\begin{itemize}

\item The value of $\alpha_{CE}$ is most sensitive to the measured fraction of He-core WDs among systems with short orbital period (below one day). 

\item If recombination energy plays a significant role, the orbital period distribution of PCEBs containing massive WDs should extend to very long periods (up to several hundred days). 

\item The secondary mass distribution for a given WD mass should reflect the IMRD. 

\end{itemize}

We are admittedly relatively far from reaching these goals. For example, we have just one observed PCEB with a massive companion (IK\,Peg). Because it might well be that the CE
efficiencies depend on the mass of the secondary star \citep{politano+weiler07-1,davisetal10-1,demarcoetal11-1}, it is not only required that we measure more orbital periods of PCEBs from
SDSS, but it is also urgent that observational surveys be extended to higher secondary masses.

\begin{acknowledgements}
MZ acknowledges support from CONICYT/FONDECYT/POSTDOCTORADO/3130559. MRS 
thanks FONDECYT (project 1100782 and 114126) and the Millennium Science Initiative, Chilean Ministry of Economy, Nucleus P10-022-F. The work of EG--B, ST, and JC was partially supported by the AGAUR, by MCINN grant AYA2011--23102, by the European Union FEDER funds, by the ESF 
EUROGENESIS project, and by the AECI grant A/023687/09. ARM acknowledges financial support from a LAMOST fellowship, from the Postdoctoral Science Foundation of China 
(grant 2013M530470), and from the Research Fund for International Young Scientists by the National Natural Science Foundation of China (grant 11350110496).
The research leading to these results has received funding from the European Research Council under the European Union's Seventh Framework Programme
(FP/2007-2013) / ERC Grant Agreement n. 320964 (WDTracer). BTG was supported in part by the UK’s Science and Technology Facilities Council (ST/I001719/1).
\end{acknowledgements}

\bibliographystyle{aa}

\end{document}